\documentclass[a4paper]{amsart}

\usepackage[margin=3.5cm]{geometry}

\usepackage{amsmath}
\usepackage[round]{natbib}
\usepackage{pxfonts}
\usepackage{mathabx}
\usepackage{natbib}

\newcommand{\quot}[1]{\ulcorner#1\urcorner}
\newcommand{\sep}{\;|\;}
\newcommand{\Var}{\mathsf{Var}\;}
\newcommand{\Lam}{\mathsf{Lam}\;}
\newcommand{\App}{\mathsf{App}\;}
\newcommand{\Term}{\mathsf{Term}}
\newcommand{\Termnf}{\mathsf{Term_{NF}}}
\newcommand{\Terma}{\mathsf{Term_{A}}}
\newcommand{\Termcps}{\mathsf{Term_{CPS}}}
\newcommand{\Termval}{\mathsf{Term_{V}}}
\newcommand{\case}[2]{\mathsf{case}\;#1\;\mathsf{of} \\ #2}
\newcommand{\interp}[1]{[\![#1]\!]}

\author{Mathieu Boespflug}
\title{From self-interpreters to normalization by evaluation}

\begin{document}
\maketitle

\begin{abstract}
  We characterize normalization by evaluation as the composition of a
  self-interpreter with a self-reducer using a special representation scheme,
  in the sense of \citet{mogensen:selfint}. We do so by deriving in a
  systematic way an untyped normalization by evaluation algorithm from a
  standard interpreter for the $\lambda$-calculus. The derived algorithm is
  not novel and indeed other published algorithms may be obtained in the same
  manner through appropriate adaptations to the representation scheme.
\end{abstract}

\section{Self-interpreters and self-reducers}
What is a self-interpreter? For the untyped $\lambda$-calculus,
\citet{mogensen:selfint} offers the following definition, given an injective
mapping $\quot\cdot$ (the {\em representation scheme}) that yields {\em
  representations} of arbitrary terms:
\[
E\;\quot M =_\beta M.
\]
That is, a self-interpreter is a term $E$ of the $\lambda$-calculus such that
when applied to the representation $\quot M$ of any term $M$, the result is a
convertible term (modulo renaming).


The $\quot\cdot$ mapping\footnote{This is an analogue of the \texttt{quote}
  special form of Scheme.} cannot of course be defined within the
$\lambda$-calculus itself, but we posit its existence as a primitive operation
of the calculus. The representation of a term is a piece of data, something
that can be manipulated, transformed and inspected within the calculus itself.
It is natural to represent data as terms in normal form, so that data may be
regarded as constant with regard to term reduction. Consider the following
grammar for terms and normal terms:
\begin{alignat*}{4}
&\Var    &\quad&\ni&\quad& x, y, z \\
&\Term   &\quad&\ni&\quad& t   &\quad&\Coloneqq x \sep \lambda x.t \sep t\;t \\
\Term \;\supset\; &\Termnf &\quad&\ni&\quad& t_n &\quad&\Coloneqq t_a \sep \lambda x.t_n \\
\Term \;\supset\; &\Terma  &\quad&\ni&\quad& t_a &\quad&\Coloneqq x \sep t_a\;t_n
\end{alignat*}
The representation scheme can be typed as $\quot\cdot : \Term \rightarrow
\Termnf$.

All manner of representation schema are possible, but
\citeauthor{mogensen:selfint} commits to a particularly simple representation
scheme, one that enables him to implement a trivially simple self-interpreter
that not only yields convertible terms from their representations, but in fact
whose weak head normal form when applied to a normal term $M$ is identical to
$M$, up to renaming of variables. Let us call this particular self-interpreter
$E_\alpha$. We have that
\[
E_\alpha\;\quot M \longrightarrow_{\mathrm{whnf}} M.
\]

\citeauthor{mogensen:selfint} goes on to define a self-reducer as a
transformation on representations:
\[
R\;\quot M =_\beta \quot{NF_M},
\]
where $NF_M$ stands for the normal form of $M$, if one exists. Equipped with
such a contraption, we can define a special kind of self-interpreter with the
additional property that all representations of terms evaluate to normal
forms. For all $M$,
\[
E_{NF}\;\quot M \hat= E_\alpha(R\;\quot M) \longrightarrow_{\mathrm{whnf}} NF_M.
\]

After a small detour towards concrete implementations of the above, we will
show how through successive transformations we may obtain an untyped
normalization by evaluation algorithm, in fact precisely the algorithm
presented in \citep{boespflug:efficientnbe}. This algorithm is but one of
several variants of untyped normalization by evaluation algorithms, and indeed
the transformations presented here can readily be adapted to derive the
algorithms of \citet{aehlig:cin} and \citet{filinski:dau}. Because the
starting points of these transformations are in fact standard definitions of
evaluators and normalizers, it is possible to obtain the correctness of the
normalization by evaluation algorithm as a corollary of the correctness of the
meaning preserving transformations we use.

\section{Implementation}

In the following, we assume a $\lambda$-calculus with matching constructs, for
convenience. Note however, that we could always reformulate the following in
the pure $\lambda$-calculus via a Church encoding.

Let us recall the representation scheme given in \citep{mogensen:selfint}:
\begin{align*}
  \quot{x} &= \Var\;x \\
  \quot{\lambda x.t} &= \Lam (\lambda x. \quot t) \\
  \quot{t_0\;t_1} &= \App \quot{t_0}\;\quot{t_1}
\end{align*}
for distinguished constructors $\Var$, $\Lam$ and $\App$. Notice how the
representation scheme uses higher order abstract syntax (HOAS). With this
representation scheme, the definition of the $E_\alpha$ self-interpreter above
can be given as
\begin{align*}
  E_\alpha\;(\Var x) &= x \\
  E_\alpha\;(\Lam t) &= t \\
  E_\alpha\;(\App t_0\;t_1) &= (E_\alpha\;t_0)\;(E_\alpha\;t_1)
\end{align*}

Given a datatype for terms, and assuming the metalanguage follows a
call-by-value evaluation strategy, an evaluator yielding weak head
normal forms using a call-by-name strategy might typically be defined
as follows:
\begin{align*}
  eval\;(\Var x) &= \Var x \\
  eval\;(\Lam t) &= \Lam t \\
  eval\;(\App t_0\;t_1) &= \case{eval\;t_0}{
    \begin{split}
      \Lam t &\rightarrow eval\;(t\;t_1) \\
      t_0' &\rightarrow \App t_0'\;(eval\;t_1)
    \end{split}
    }
\end{align*}
Now a single change to the above gives us the definition of a normalizer:
\begin{align*}
  norm\;(\Var x) &= \Var x \\
  norm\;(\Lam t) &= \Lam (\lambda x. norm\;(t\;x)) \\
  norm\;(\App t_0\;t_1) &= \case{norm\;t_0}{
    \begin{split}
      \Lam t &\rightarrow t\;t_1 \\
      t_0' &\rightarrow \App t_0'\;(norm\;t_1)
    \end{split}
    }
\end{align*}
Whereas the evaluator does not traverse the boundary of a binder to fetch
redexes beneath it (an abstraction is a value), a normalizer on the other hand
does, so we move the recursive call to $norm$ inside the abstraction. There
are no more recursive calls in the definition of $norm$ than there are in the
definition of $eval$ --- one of them merely changed place.

A slight modification gives a normalizer following the call-by-value strategy.
Namely, we normalize the argument before applying any abstraction to it.
\begin{align*}
  norm\;(\Var x) &= \Var x \\
  norm\;(\Lam t) &= \Lam (\lambda x. norm\;(t\;x)) \\
  norm\;(\App t_0\;t_1) &= \case{norm\;t_0}{
    \begin{split}
      \Lam t &\rightarrow t\;(norm\;t_1) \\
      t_0' &\rightarrow \App t_0'\;(norm\;t_1)
    \end{split}
    }
\end{align*}
Notice that the call-by-value and call-by-name normalizers just defined both
fit the bill as self-reducers: they take representations of terms to
representations in normal form. That is, if one is willing to forgo the
incompleteness of the call-by-value normalizer: there are terms for which
there exists a normal form that may never be reached using a call-by-value
strategy.

The need to define two separate normalizers for two different normalization
strategies is rather unfortunate. But if we restrict the input terms to a
certain shape, namely those terms in continuation passing style (CPS), then
there is only one normalization strategy possible, because at any given step
in the computation, the only possible redex outside of any abstraction is to
be found at the head of the term, if any. Hence only one normalizer need be
built. But more importantly, it will justify a change in the representation
scheme allowing for a much more efficient self-reducer than is possible with
the current representation scheme.

\section{Shifting representation scheme}

A CPS transformation takes terms in $\Term$ to terms in $\Termcps$, a language
generated by the following grammar.
\begin{alignat*}{4}
\Term \;\supset\; &\Termval &\quad&\ni&\quad& v &\quad&\Coloneqq x
\sep \lambda x.t_c \\
\Term \;\supset\; &\Termcps &\quad&\ni&\quad& t_c &\quad&\Coloneqq v
\sep v\;v
\end{alignat*}
By modifying the representation scheme to return representations in
continuation passing style instead, we effectively encode the reduction
strategy of the normalizer into the scheme itself\footnote{Alternatively, we
  could leave the representation scheme untouched and have the CPS transform
  as a function from representations to representations, which we intercalate
  between quoting and normalizing.}. But on the flip side, we get a simpler
definition for the normalizer:
\begin{align*}
  norm\;(\Var x) &= \Var x \\
  norm\;(\Lam t) &= \Lam (\lambda x. norm\;(t\;x)) \\
  norm\;(\App (\Lam t_0)\;t_1) &= norm\;(t_0\;t_1) \\
  norm\;(\App (\Var x)\;t_1) &= \App (\Var x)\;(norm\;t_1)
\end{align*}
This definition is a mere specialization of the above definitions to terms in
CPS, noting that such terms do not contain nested applications. We can break
out the interpretation of application nodes to an auxiliary $app$ function,
using the fact that $\App (\Var x)\;(norm\;t_1) \equiv norm\;(\App (\Var
x)\;t_1)$ for all $x$, $t_1$:
\begin{align*}
  norm\;(\Var x) &= \Var x \\
  norm\;(\Lam t) &= \Lam (\lambda x. norm\;(t\;x)) \\
  norm\;(\App t_0\;t_1) &= norm\;(app\;t_0\;t_1) \\
  \\
  app\;(\Lam t_0)\;t_1 &= t_0\;t_1 \\
  app\;(\Var x)\;t_1 &= \App (\Var x)\;t_1
\end{align*}

\section{From here to there: normalization by evaluation}

Normalization by evaluation corresponds to interpreting terms as their
representation yet identifying entire classes of terms with unique
representations. In our setting, this is achieved by inlining the third clause
of $norm$ in place of every occurrence of an $\mathsf{App}$ node in the
representation of the term to normalize. As an optimization, we can inline
only the call to $app$, not the recursive call to $norm$, yielding the
following final definition for $norm$,
\begin{align*}
  norm\;(\Var x) &= \Var x \\
  norm\;(\Lam t) &= \Lam (\lambda x. norm\;(t\;x)) \\
  norm\;(\App t_0\;t_1) &= \App t_0\;(norm\;t_1)
\end{align*}
where we use the fact that after inlining of $app$ all $\mathsf{App}$ nodes
that remain are applicative forms, i.e.\ bearing a variable as left branch.
The associated interpretation function reads as follows:
\begin{align*}
\interp{x} &= \Var x \\
\interp{\lambda x. t} &= \Lam (\lambda x. \interp t) \\
\interp{t_0\;t_1} &= app\;\interp{t_0}\;\interp{t_1}
\end{align*}
This interpretation function can be read as a particular quote operation where
terms are identified modulo weak head normal forms.

Normalization by evaluation is
\begin{align*}
  nbe\;t &= E_\alpha\;(norm\;\interp{t})
\end{align*}

\section{Conclusion}

We have gone from a self-reducer to an equivalent self-reducer on
representations in CPS. By inlining part of the obtained self-reducer into the
representation of terms, we further obtained a preexisting normalization by
evaluation algorithm (which also works for terms in direct style). In short,
for terms in CPS, untyped normalization by evaluation is the composition of a
self-reducer with an appropriate quote operation.

We have derived a more efficient self-reducer than the one presented in
\citep{mogensen:selfint}. This derivation also serves as a road map for an
alternative and simple proof of correctness of untyped normalization by
evaluation for terms in CPS.

\section*{Acknowledgements}

The author wishes to dedicate this short note to Olivier Danvy. Once upon a
studious afternoon, its essence fell out rather immediately from the following
question when pondering an unrelated problem: ``How would Olivier Danvy
think?''.

\bibliographystyle{plainnat}
\bibliography{references}

\begin{thebibliography}{4}
\providecommand{\natexlab}[1]{#1}
\providecommand{\url}[1]{\texttt{#1}}
\expandafter\ifx\csname urlstyle\endcsname\relax
  \providecommand{\doi}[1]{doi: #1}\else
  \providecommand{\doi}{doi: \begingroup \urlstyle{rm}\Url}\fi

\bibitem[Aehlig et~al.(2008)Aehlig, Haftmann, and Nipkow]{aehlig:cin}
K.~Aehlig, F.~Haftmann, and T.~Nipkow.
\newblock {A Compiled Implementation of Normalization by Evaluation}.
\newblock \emph{Theorem Proving in Higher Order Logics (TPHOLs 2008), Lecture
  Notes in Computer Science. Springer-Verlag}, 2008.

\bibitem[Boespflug(2009)]{boespflug:efficientnbe}
Mathieu Boespflug.
\newblock Efficient normalization by evaluation.
\newblock In \emph{Informal proceedings of the 2009 Workshop on Normalization
  by Evaluation}, pages 29--34, August 2009.

\bibitem[Filinski and Rohde(2004)]{filinski:dau}
A.~Filinski and H.K. Rohde.
\newblock {A denotational account of untyped normalization by evaluation}.
\newblock \emph{Lecture notes in computer science}, pages 167--181, 2004.

\bibitem[Mogensen(1992)]{mogensen:selfint}
T.A. Mogensen.
\newblock {Efficient self-interpretation in lambda calculus}.
\newblock \emph{Journal of Functional Programming}, 2\penalty0 (03):\penalty0
  345--364, 1992.

\end{thebibliography}

\end{document}